\documentclass{elsart}
\usepackage{amsmath}
\usepackage{graphicx}
\newcommand{\NOT}{\textsc{not}}
\newcommand{\CNOT}{controlled-\NOT}

\begin{document}

\begin{frontmatter}
\title{Suppressing Weak Ising Couplings: Tailored Gates for Quantum Computation}
\author{Jonathan A. Jones}
\address{Centre for Quantum Computation, Clarendon Laboratory, Parks Road, Oxford OX1 3PU, United Kingdom.}
\corauth{Email address: jonathan.jones@qubit.org}

\begin{abstract}
I describe the use of techniques based on composite rotations to
develop controlled phase gates in which the effects of weak Ising
couplings are suppressed.  A tailored composite phase gate is
described which both suppresses weak couplings and is relatively
insensitive to systematic errors in the size of strong couplings.
\end{abstract}

\end{frontmatter}

Quantum computers \cite{bennett00} are extremely vulnerable to the
effects of errors, and there has been considerable interest in
correcting random errors arising from decoherence processes
\cite{shor95,steane96,steane99} and, more recently, in stabilising
logic gates against systematic errors, which arise from
reproducible imperfections in the system used to implement quantum
computations.  In the context of Nuclear Magnetic Resonance (NMR)
quantum computation \cite{cory96,cory97,gershenfeld97,jones01a}
systematic errors can be reduced using composite rotations
(composite pulses) \cite{ernst87,freeman97b,levitt86,wimperis94},
which have been applied to both single qubit gates
\cite{cummins00,cummins03} and the two qubit controlled
phase-shift gate \cite{jones03a,jones03b}.

Here I consider a related but distinct problem: the design of
composite phase gates which automatically suppress evolution under
small Ising couplings.  This is useful, as many proposed
implementations involve extended networks of Ising couplings.
Typically in such systems the nearest neighbour interactions are
large, and form the basis of two qubit gates, while next-nearest
neighbour interactions are much weaker.  Frequently such
interactions are too weak to be useful, but too strong to simply
be ignored: it is, therefore, useful to develop sequences which
are capable of distinguishing between small and large couplings,
effectively suppressing small couplings.  Ideally such sequences
should also be robust to small errors in the size of the large
couplings.

As described previously \cite{jones03a,jones03b} it is
straightforward to convert a robust single qubit gate into a
robust controlled phase-shift gate; as single qubit gates are much
easier to study it makes sense to adopt this approach.  A
composite rotation comprises a series of rotations applied in
sequence, such that the overall effect in the absence of errors is
the desired rotation, while in the presence of \textit{systematic}
errors the effect of these errors largely cancels out.  With some
approaches to designing composite rotations it can be difficult to
ensure that both requirements are simultaneously true, but this
problem can be side-stepped using a method due to Wimperis
\cite{wimperis94}, in which the composite rotation is constructed
by combining a simple rotation with a complex series of additional
rotations, which in the absence of errors does nothing. This
approach guarantees that the composite rotation will behave
correctly in the absence of errors, while allowing error tolerance
to be developed by tweaking the do-nothing sequence.  Note that
although these composite rotations are developed in the context of
NMR they are applicable to other implementations of quantum
computation based on Ising couplings
\cite{lloyd93,ioffe99,cirac00,briegel01,raussendorf01}.

Single qubit gates are transformations in SU(2), and so can be
conveniently represented by quaternions \cite{cummins03}.  The
composite quaternion for a composite rotation can be obtained by
multiplying the quaternions for the individual rotations, and then
expanding the result as a Maclaurin series in the fractional
error.  As an example consider the BB1 sequence
\cite{wimperis94,cummins03} for reducing the effects of errors in
rotation rates (traditionally called pulse length errors in NMR):
the simple pulse $\theta_x$ is replaced by the composite pulse
sequence
\begin{equation}
(\theta/2)_x 180_{\phi_1} 360_{\phi_2} 180_{\phi_1} (\theta/2)_x
\end{equation}
where the phase angles $\phi_1$ and $\phi_2$ are chosen to make
the rotation error tolerant.  If the rotation errors are
parameterised using the fractional error in the rotation rate,
$g$, then the simple rotation shows first order errors in $g$, but
these can be removed by choosing $\phi_2=3\phi_1$ and
\begin{equation}
\phi_1=\pm\arccos\left(-\frac{\theta}{4\pi}\right).
\label{eq:phi1}
\end{equation}
Remarkably this approach also removes the errors which are second
order in $g$.

The BB1 approach can easily be adapted to tackle the new problem:
developing a composite rotation which has only minimal effect when
the rotation rate is very small.  (When converted to its two qubit
equivalent, this will create a controlled phase-shift gate which
effectively suppresses evolution under small Ising couplings).
This can be achieved by comparing the composite quaternion with
the null quaternion
\begin{equation}
\mathsf{q_0}=\{1,\{0,0,0\}\}
\end{equation}
and then replacing the Maclaurin series expansion with a Taylor
series expansion around the point $g=-1$. As observed previously
\cite{jones03b}, the $z$-component of the composite quaternion is
automatically correct, while the first order $y$-component can be
removed by choosing $\phi_2=-\phi_1$.  Finally $\phi_1$ must be
chosen to zero the first order $x$-component, and is once again
given by Eq.~\ref{eq:phi1}.

\begin{figure*}
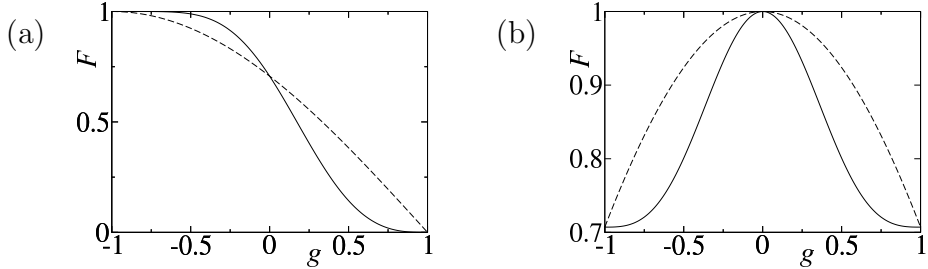

\raisebox{85pt}{(a)}\quad\includegraphics*[scale=0.2]{fig1a.eps}\qquad
\raisebox{85pt}{(b)}\quad\includegraphics*[scale=0.2]{fig1b.eps}
\caption{Fidelity $F$ of simple (dashed line) and selective NB1
(solid line) $90^\circ$ rotations as a function of the fractional
error in the rotation rate $g$.  Fidelities are measured (a)
against an ideal $0^\circ$ rotation (the identity operation), and
(b) against a $90^\circ$ rotation. An ideal composite rotation
would show broad plateaus of height $1$ at $g=-1$ in plot (a) and
at $g=0$ in plot (b).} \label{fig:NB1}
\end{figure*}

The results of this composite rotation (which will be called NB1
for reasons that will become clear later) for the case
$\theta=90^\circ$ are shown in Fig.~\ref{fig:NB1}.  This shows the
fidelity $F$ of simple and composite rotations as a function of
the fractional error in the rotation rate $g$.  Note that for
single qubit gates the propagator and quaternion fidelities are
the same \cite{cummins03,jones03a,jones03b}. An ideal composite
rotation would show broad plateaus of height $1$ at $g=-1$ in plot
(a), indicating effective suppression of small couplings, and at
$g=0$ in plot (b), indicating robust evolution under large
couplings. Clearly the NB1 composite rotation is effective at
suppressing evolution around $g=-1$, but it achieves this at the
expense of reducing the robustness of the sequence in the
low-error region, around $g=0$. This is not particularly
surprising, and becomes even less so on the discovery that this
composite rotation is in fact essentially identical to the first
narrowband excitation family (NB1) described by Wimperis
\cite{wimperis94}.  (The only difference is that the new version
has been time symmetrised, by placing the correction pulses half
way through the main pulse, rather than at the start).

\begin{figure*}
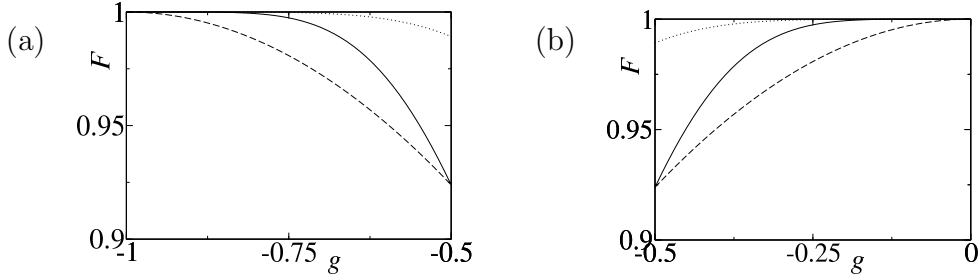

\raisebox{85pt}{(a)}\quad\includegraphics*[scale=0.2]{fig2a.eps}\qquad
\raisebox{85pt}{(b)}\quad\includegraphics*[scale=0.2]{fig2b.eps}
\caption{Fidelity of simple (dashed line), PB1 (solid line), and
NB1 or BB1 (dotted line) $90^\circ$ rotations as a function of the
fractional error in the rotation rate $g$: (a) fidelity measured
against an ideal $0^\circ$ rotation (the identity operation), with
the dotted line showing the NB1 sequence; (b) fidelity measured
against a $90^\circ$ rotation, with the dotted line showing the
BB1 sequence.  Note that the horizontal axes differ in the two
plots.} \label{fig:PB1}
\end{figure*}

As two interesting composite rotations (BB1 and NB1) are in fact
rediscoveries of results by Wimperis, it seems sensible to examine
his other results to see what else might be found.  He describes
\cite{wimperis94} two other families of broadband and narrow band
pulses, BB2 and NB2, but these are not suitable for quantum
computing.  He also describes two families of ``passband''
composite rotations, PB1 and PB2, and remarkably PB1 is exactly
the sequence we are seeking (PB2 is not suitable for quantum
computing). The time symmetrised version takes the form
\begin{equation}
(\theta/2)_x 360_{\phi_1} 720_{\phi_2} 360_{\phi_1} (\theta/2)_x
\end{equation}
with $\phi_2=-\phi_1$ and
\begin{equation}
\phi_1=\pm\arccos\left(-\frac{\theta}{8\pi}\right). \label{eq:PB1}
\end{equation}
The performance of the PB1 sequence is shown in
Fig.~\ref{fig:PB1}.  This sequence is in some sense a compromise
between NB1 and BB1: it performs worse than NB1 near $g=0$ and
worse than BB1 near $g=-1$, but outperforms a simple pulse in both
domains, exactly as desired.

\begin{figure}
\begin{center}
\includegraphics{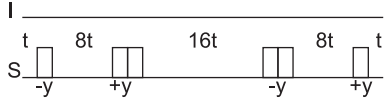}
\end{center}
\caption{Pulse sequence for an Ising gate to implement a \CNOT\
gate which both suppresses small couplings and is robust to small
errors in coupling sterngths. Boxes correspond to single qubit
rotations with rotation angles of
$\phi=\arccos(-1/16)\approx93.6^\circ$ applied along the $\pm y$
axes as indicated; time periods correspond to free evolution under
the Ising coupling, $\pi J\,2I_zS_z$ for multiples of the time
$t=1/4J$.  The naive Ising gate corresponds to free evolution for
a time $2t$.} \label{fig:pulses}
\end{figure}

Finally I return to the problem of converting a single qubit PB1
composite rotation into the desired two qubit controlled
phase-shift gate, which both suppresses small couplings and is
robust to small errors in coupling strengths.  The controlled
phase-shift gate is equivalent to the Ising coupling gate
\cite{jones01a}, and a robust Ising gate can be built as described
previously \cite{jones03a}.  After combining and cancelling
extraneous pulses, the final sequence for the case $\theta=\pi/2$
(which forms the basis of the \CNOT\ gate) is shown in
Fig.~\ref{fig:pulses}. For two qubit gates the natural definition
of the fidelity is given by the propagator fidelity
\cite{jones03a}, but for two qubit gates of this kind, which are
essentially equivalent to single qubit gates, the propagator
fidelity is once again equivalent to the quaternion fidelity
\cite{jones03b}.

Following previous work \cite{jones03a}, I assume that an
infidelity ($1-F$) of $10^{-6}$ can be tolerated in a quantum
logic gate.  A useful measure of the practicality of an Ising
coupling gate is then provided by the maximum fractional error
$\epsilon$ in the coupling constant $J$ before this tolerance is
exceeded \cite{jones03b}.  Another useful measure, indicating the
ability of the gate to suppress small couplings, is the largest
fractional coupling $\delta$ which cannot simply be neglected.  On
these measures the simple Ising gate does not perform well: it
requires that $J$ be controlled to better than 0.2\%
($\epsilon<0.0018$) and the limit on small couplings is the same
($\delta<0.0018$).  The BB1 gate corrects well for small errors in
$J$, permitting errors of up to 10\% ($\epsilon<0.1015$), but is
even more sensitive to small couplings, requiring $\delta<0.0009$.
As might be guessed from the symmetry of the two plots in
Fig.~\ref{fig:PB1} the performance of NB1 is exactly the opposite,
allowing small couplings to be effectively suppressed
($\delta<0.1015$), but at the cost of increased sensitivity to
errors ($\epsilon<0.0009$).  Finally the PB1 sequence is a
compromise between NB1 and BB1, with $\delta<0.0648$ and
$\epsilon<0.0648$.

For most purposes the PB1 sequence provides the most practical
Ising gate currently known: at an infidelity of $10^{-6}$ it can
tolerate errors of more than 6\% in coupling constants, but it
\textit{also} permits small couplings, with strengths up to 6\% of
the coupling being used to implement a gate, to be neglected.  If
even higher fidelities are required then these bounds will of
course be reduced, but the relative improvement provided by the
PB1 sequence will be even greater.

%\begin{acknowledgments}
I thank S. Benjamin for helpful conversations and the UK EPSRC for
financial support.
%\end{acknowledgments}

\end{document}